\documentclass[aps,prb,twocolumn,superscriptaddress]{revtex4}

\usepackage[pdftex]{graphicx}
\usepackage[latin1]{inputenc}

\begin{document}

\title{Spatial motion of the magnetic avalanches associated to the CO-AFM to CD-FM transition in La$_{0.225}$Pr$_{0.40}$Ca$_{0.375}$MnO$_{3}$
manganite}
\author{A. \surname{Hern\'andez-M\'inguez}}
\author{F. \surname{Maci\`a}}
\author{J. M. \surname{Hernandez}}
\author{G. \surname{Abril}}
\author{A. \surname{Garc\'ia-Santiago}}
\author{J. \surname{Tejada}}
\affiliation{Departament de Física Fonamental, Facultat de Física,
Universitat de
Barcelona\\
Avda. Diagonal 647, Planta 4, Edifici nou, 08028 Barcelona, Spain}
\author{F. \surname{Parisi}}
\affiliation{Departamento de Física. Comisión Nacional de Energía
Atómica and Escuela de Ciencia y Tecnología, UNSAM, Av. Gral Paz
1499 (1650) San Martin. Buenos Aires. Argentina.}

\date{\today}

\begin{abstract}
Very fast magnetic avalanches in (La, Pr)-based manganites are the
signature of a phase transition from an insulating blocked
charge-ordered (CO-AFM) state to a charge delocalized
ferromagnetic (CD-FM) state. We report here the experimental
observation that this transition does not occur neither
simultaneously nor randomly in the whole sample but there is
instead a spatial propagation with a velocity of the order of tens
of m/s. Our results show that avalanches are originated in the
inside of the sample, move to the outside and occur at values of
the applied magnetic field that depend on the CD-FM fraction in
the sample. Moreover, a change in the gradient of the magnetic
field along the sample shifts the point where the avalanches are
ignited.
\end{abstract}

\maketitle

The study of the coexistence of the two well segregated
charge-delocalized ferromagnetic (CD-FM) and charge-ordered
antiferromagnetic (CO-AFM) phases in manganites, as well as the
influence of the applied magnetic field on the occurrence of the
magnetic avalanches accompanying the transition from the CO-AFM
phase to the CD-FM phase has nowadays a resurgent interest driven
by the unusual (but not uncommon) interplay between static and
dynamic properties of the phase separated (PS)
state.\cite{MathurNatMat,Wu,DagottoScience} Phase separation can
appear near the boundary of a first order phase transition
(FOPT).\cite{Dagotto} In disordered-free magnetic systems the FOPT
occurs at a sharply-defined line in the magnetic field-temperature
$(H-T)$ space. Quenched impurities can lead, under certain
circumstances, to the spread of the local transition temperatures
(where local means over length scales of the order of the
correlation length) leading to the appearance of clustered states
with the consequent rounding of the FOPT.\cite{Imry} The phase
separation observed in manganites seems to belong to this last
class of systems, being the PS state the true equilibrium state of
the system in the vicinity of the FOPT,\cite{Moreo&Mayr,Levy} and
not just the result of supercooling or superheating of the system.
It has been shown\cite{Ghiev2} that the response of the system
within the PS state is governed by a hierarchical cooperative
dynamics, with state-dependent energy barriers which diverge as
the system approaches equilibrium. This slow growing dynamics,
which resembles that of glass-like systems, has given place to the
construction of phase diagrams of manganites focused on the
dynamic properties of the PS state, with regions of the phase
diagrams named as "frozen" or "dynamic" PS\cite{Ghiev2} "strain
glass" or "strain liquid",\cite{Sharma} and the determination of
state-dependent blocking or freezing
temperatures.\cite{Ghiev2,Sacanell} These kinds of processes are
not exclusive of PS manganites, but are also observed in other
systems displaying magnetic FOPT.\cite{Kumar,Chatto}

Direct evidences of micrometer phase coexistence in manganites at
intermediate temperatures was obtained through electron
microscopy\cite{UeharaNat,Murakami} and magnetic force microscopy
(MFM).\cite{Zhang} A major goal in the description of the low
temperature behavior of the PS manganite
La$_{5/8-y}$Pr$_y$Ca$_{3/8}$MnO$_3$ [LPCM($y$), $y=3/8$] was
achieved recently by Wu \textit{et al.}.\cite{Wu} Using a MFM
technique they were able to "photograph" the isothermal evolution
of the system as $H$ varies (a metamagnetic transition), and the
evolution at fixed $H$ when $T$ increases (the "glass" transition)
at the micrometer scale, giving direct evidence of the growing
process of the CD-FM phase against the CO-AFM. They conclude, in
agreement with the thermodynamic consideration presented in Ref.
\onlinecite{Sacanell}, that the low temperature state accessible
after cooling the sample in zero magnetic field could be
associated with the supercooled state of the CO-AFM to CD-FM FOPT,
the kinetics of the transformation being influenced by
accommodation strain between the coexisting phases. The results
presented in Ref. \onlinecite{Wu} were obtained at temperatures
above 6 K. Below this temperature the metamagnetic transition
between the frozen CO-AFM state and the equilibrium CD-FM phase
does not occur in a continuous way,\cite{Ghiev1,Schiffer} at least
in the time scale of the MFM experiments (seconds), but it occurs
sharply at fields above 2 T, presumably in milliseconds.

\begin{figure}
\includegraphics[width=\columnwidth]{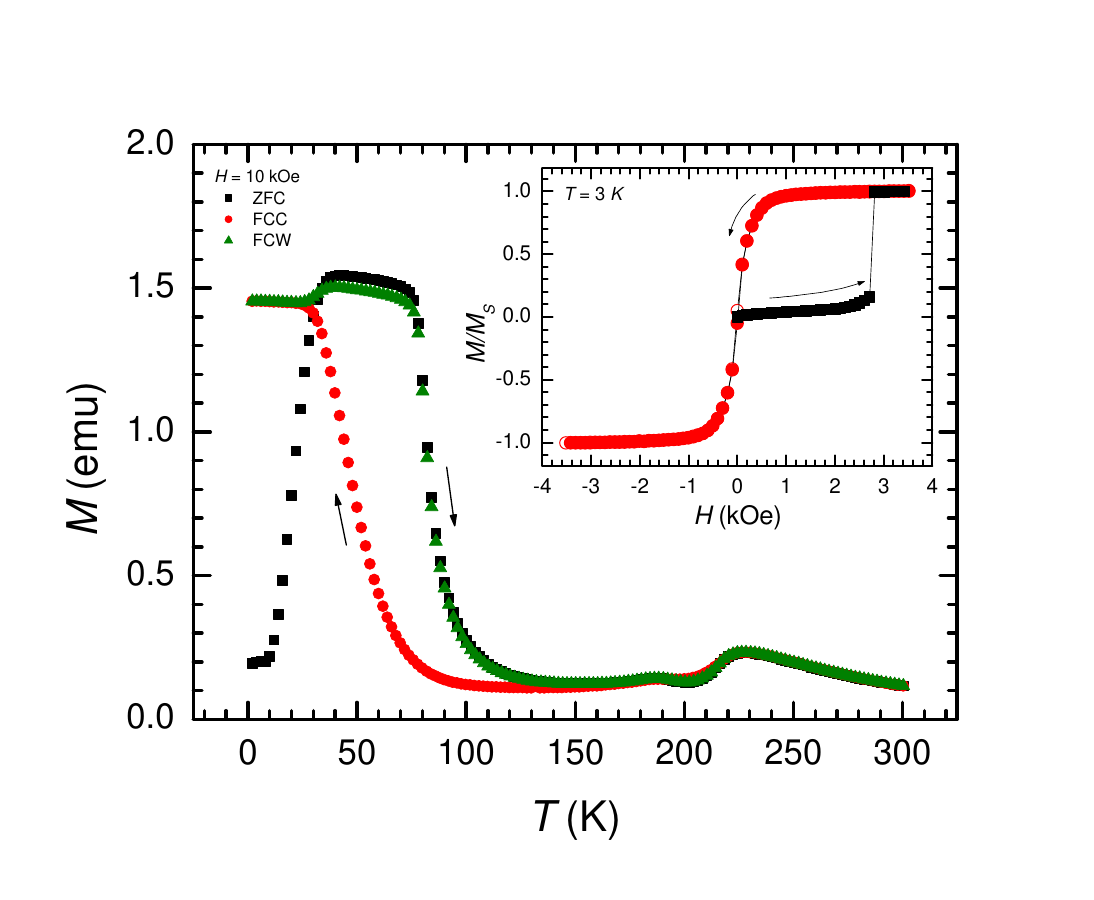}
\caption{Temperature dependence of the zero-field cooled
(squares), field-cooled on cooling (circles) and field-cooled on
warming (triangles) magnetization of
La$_{0.225}$Pr$_{0.4}$Ca$_{0.375}$MnO$_3$, measured for a magnetic
field of 10 kOe. The inset shows the field dependence of the
magnetization with respect to the saturation value, $M_s$, at 3 K.
Squares correspond to the first magnetization curve, whereas
circles show the magnetic measurements from 35 kOe to -35 kOe and
back to 35 kOe.} \label{fig:FC}
\end{figure}

In this work we present the study of the dynamics of the CO-AFM to
the CD-FM FOPT in LPCM(0.4) below 6 K, and its spatial evolution.
A polycrystalline sample of composition
La$_{0.225}$Pr$_{0.40}$Ca$_{0.375}$MnO$_3$ and dimensions $2\times
2\times 5$ mm$^3$ was used. All magnetic measurements were
performed placing the sample inside a commercial Superconducting
Quantum Interference Device magnetometer. The sample was first
characterized by measuring the temperature and magnetic field
dependences of the magnetization. Figure \ref{fig:FC} shows that,
as $T$ decreases, the sample experiences a transition from a
paramagnetic to a CO-AFM state at $T_{CO}=220$ K. A short peak at
190 K reveals the formation of small ferromagnetic clusters before
a more robust ferromagnetic transition appears at a lower
temperature ($T_C=70$ K on cooling, $T_C=90$ K on warming).
Finally, at $T_B=25$ K the blocking temperature associated to the
energy barriers between the two phases is clearly seen. The inset
in Fig. \ref{fig:FC} shows the isothermal magnetization curve
measured at 3 K. At the first magnetization curve (squares) a
magnetic avalanche was detected at 28 kOe, being the final
magnetization the saturation value, $M_s$, which corresponds to
the case of having only the ferromagnetic phase. We then kept the
temperature constant and measured, immediately after, a $M(H)$
going from 35 kOe down to -35 kOe and back to 35 kOe. This
hysteresis cycle shows that the ferromagnetic state has neither
remanence nor coercivity and, consequently, there are not time
dependent phenomena. This point has been further verified by
measuring the constancy of the magnetization after fast changes of
the magnetic field. All these results are in full agreement with
those previously reported for similar PS manganites.\cite{Ghiev1}

\begin{figure}
\includegraphics[width=\columnwidth]{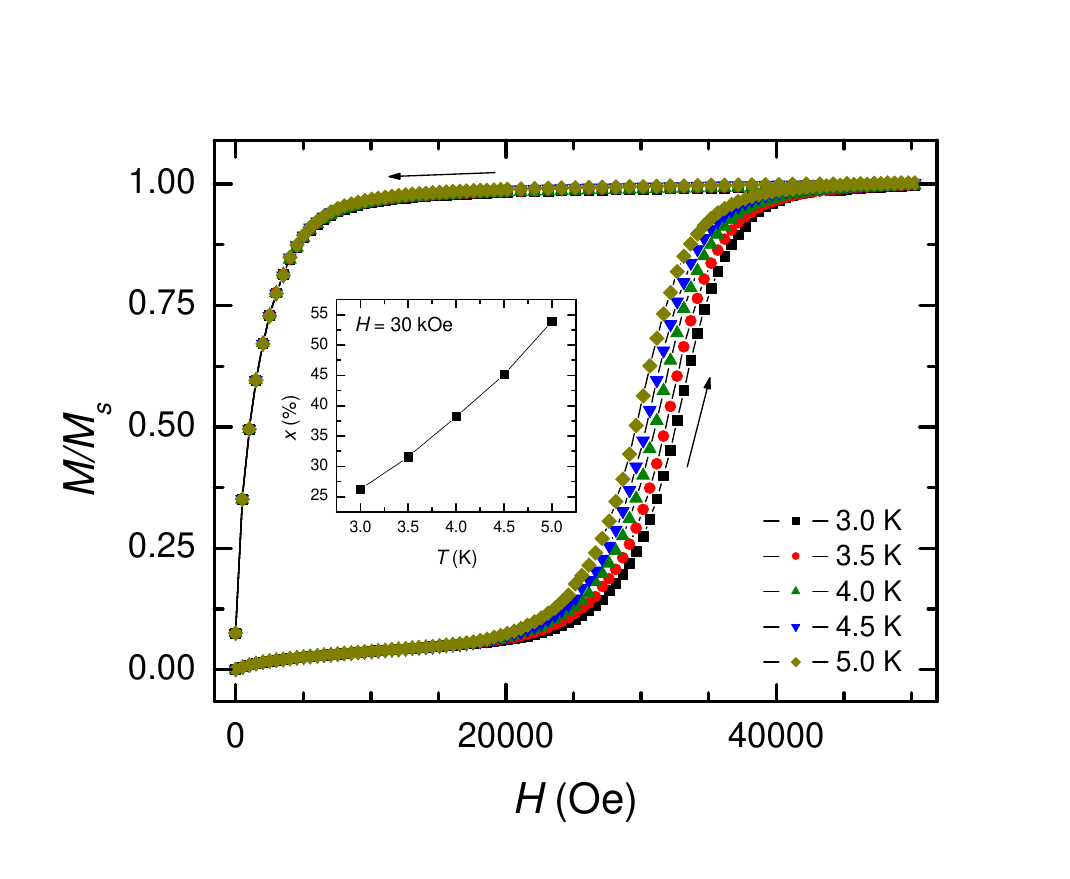}
\caption{Isothermal magnetization curves registered at several
temperatures between 3 K and 5 K after zero field cooling the
sample. The inset shows the temperature dependence of the
ferromagnetic phase fraction at the first magnetization curve,
$x$, for an applied magnetic field of 30 kOe.}\label{fig:MH}
\end{figure}

In Fig. \ref{fig:MH} we show isothermal $M(H)$ curves registered
at several temperatures comprised between 3 and 5 K after
zero-field cooling the sample. Magnetic avalanches are turned on
due to the interplay between the local increase of the FM fraction
and the heat released in this microscopic
transformation.\cite{Ghiev1} To avoid the ignition of these
avalanches, we increased the number of points that were measured
at the same interval of applied magnetic field with respect to the
hysteresis curve at the inset of Fig. \ref{fig:FC}. As each point
takes several seconds to be measured, this reduces the average
sweep rate of the applied magnetic field, allowing the sample to
thermalize and making therefore possible to register the $M$
values of the PS state at many magnetic fields and temperatures
where otherwise the sample would have already transited to the
CD-FM state. The data of Fig. \ref{fig:MH} show that first
magnetization curves for field values smaller than 20 kOe
coincide, indicating that the FM phase fraction, $x$, is nearly
the same at all temperatures between 3 and 5 K. The value of $x$
through out the first magnetization curve can be estimated, at
each temperature and magnetic field, from the ratio between $M$
and $M_s$. This definition is no longer valid once $M_s$ has been
achieved, due to the fact that, from then on, all the sample is in
the CD-FM state as long as it is not heated and zero-field cooled
again. According to this definition, a value of $x$ smaller than
8\% was estimated at 20 kOe. For magnetic fields larger than 20
kOe, there is, however, a separation between the different first
magnetization curves indicating that the FM phase fraction depends
on both the temperature and the magnetic field. Time dependent
phenomena occur in this regime of values of $T$ and $H$ indicating
that the phase separation is a dynamic process which may be
characterized by a barrier height $U(H, T)$ that separates the
CD-FM and the CO-AFM states. The changes in $x$ are therefore due
to thermal transitions above this barrier height. The temperature
dependence of $x$ for 30 kOe is given in the inset of Fig.
\ref{fig:MH}. These values of $x$ correspond, therefore, to the
blocked phase fraction compatible with the time scale of the
experiment, that in this case is of the order of seconds.

\begin{figure}
\includegraphics[width=\columnwidth]{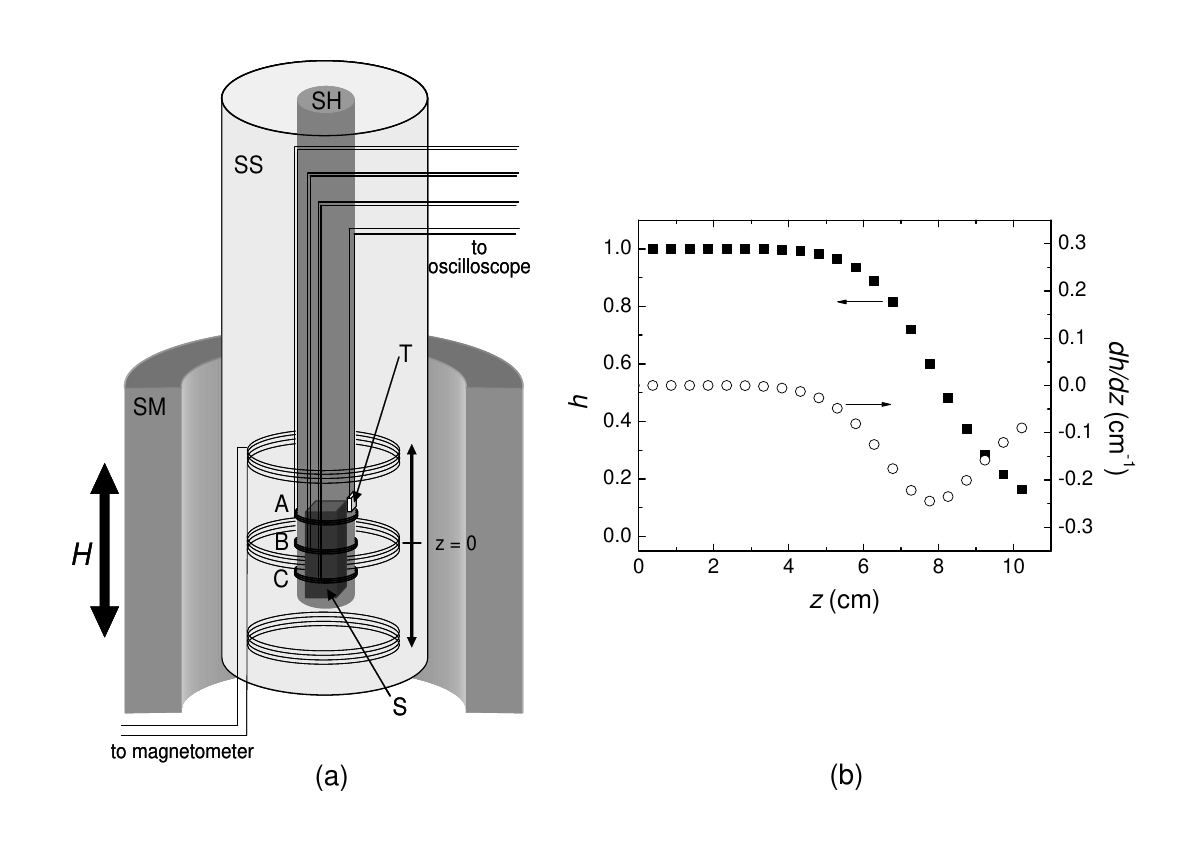}
\caption{Experimental setup. (a) The sample (S) is mounted on a
sample holder (SH) with three detection coils (A, B and C) and one
thermometer (T). This ensemble is introduced in the sample space
(SS) of a commercial magnetometer and centered in the
superconducting magnet (SM) that applies the magnetic field. The
center of the SM is taken as $z=0$. (b) Reduced applied magnetic
field (solid squares), $h(z)=H(z)/H_{max}$ ($H_{max}=H(z=0)$), and
the numerical derivative of such magnitude (open circles),
d$h$/d$z$($z$).} \label{fig:setup}
\end{figure}

In the next we will describe the experiments and results when
local magnetic measurements are used to study the magnetic
avalanches.\cite{Ahernami3} Figure \ref{fig:setup}a shows the
experimental setup. The sample was placed inside a plastic tube
and three coils of 4 turns each were wound around the tube, one in
the center (coil B) and two at the two edges of the sample (coils
A and C, respectively). In Fig. \ref{fig:setup}b we show the
spatial dependence of the reduced magnetic field,
$h=H(z)/H_{max}$, defined as the ratio between the applied
magnetic field value at the position $z$, $H(z)$, and the nominal
value of the magnetic field applied at the center of the
superconducting magnet, $H_{max}$. By placing the sample in
different positions along the $z$ axis of the cryostat we may
change the field gradient acting along one of the dimensions, as
it can be seen in Fig. \ref{fig:setup}b from the curve of the
numerical derivative of the reduced magnetic field, d$h$/d$z$. In
these experiments we placed the sample so that the so-called $z$
direction corresponded to the longest dimension of the sample (see
Fig. \ref{fig:setup}a).

\begin{figure}
\includegraphics[width=\columnwidth]{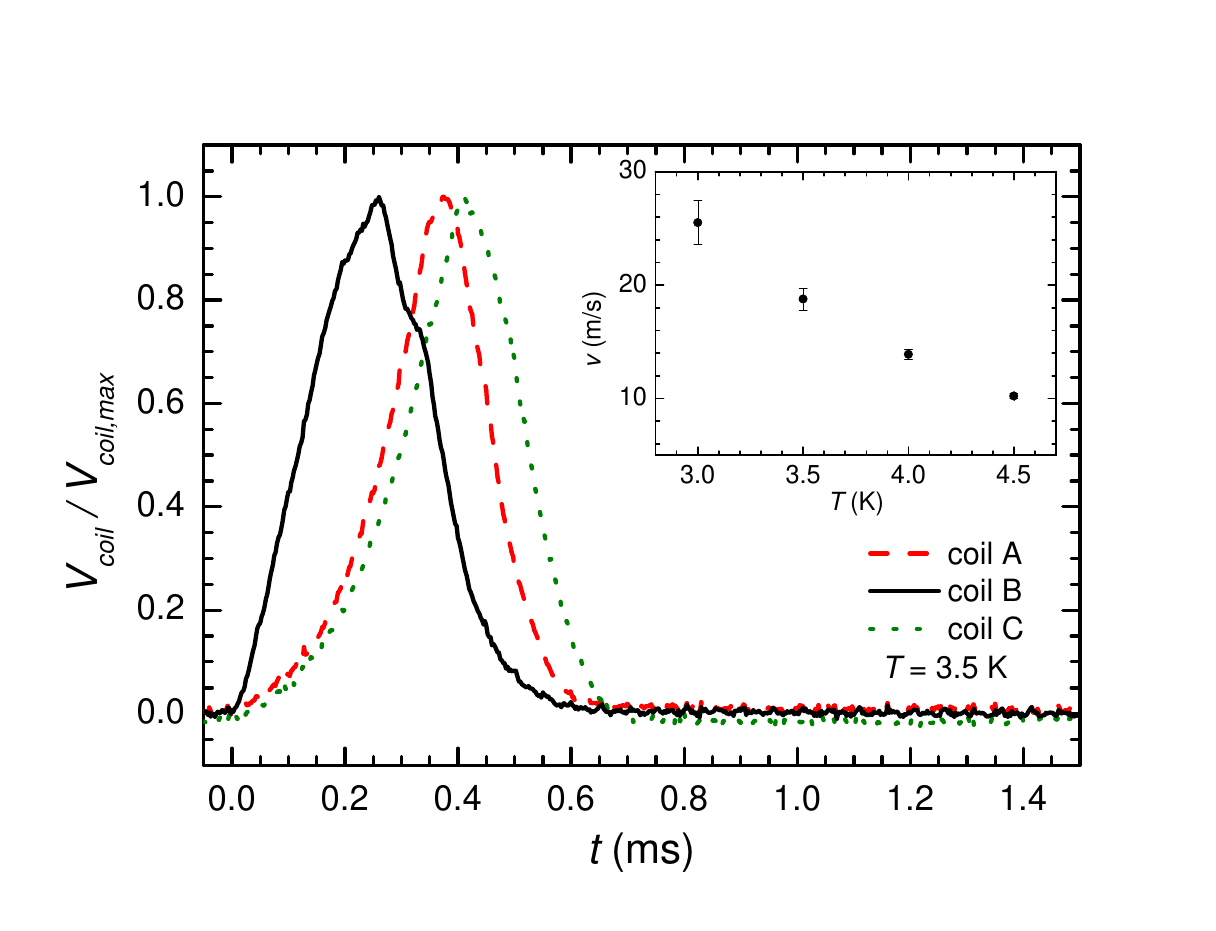}
\caption{Time variation of the magnetization as detected by coils
A (dashed line), B (solid line) and C (dotted line) at 3.5 K.
These three signals have been normalized with respect to their
maximum values. The inset shows the propagation velocity of the
phase transformation front along the sample as a function of
temperature.} \label{fig:coils}
\end{figure}

Figure \ref{fig:coils} shows the voltages detected by the three
coils, which are proportional to the variation of magnetization as
a function of time, d$M$/d$t$. $t=0$ corresponds to the instant
when the avalanche does start and so the three coils detect a
non-zero voltage value. These experiments were performed at
temperatures comprised between 3 and 5 K, when the sample was
centered at $z=0$. As we could only monitorize two signals
simultaneously, we repeated the experiment at each temperature
several times, registering two different coils each time. The
results turned to be highly reproducible and deterministic in this
temperature range, so we used the signal detected by coil B to
synchronize the signals of all three coils. From Fig.
\ref{fig:coils} we conclude that the ignition process of the
avalanche occurs systematically in the interior of the sample and
then moves to the outside. The velocity of propagation of the
transformation front can be roughly estimated as $v=L/\Delta t$,
where $L$ is the length of the sample, $\Delta t$ is defined as
$\Delta t=(t_A-t_B)+(t_C-t_B)$, and $t_A$, $t_B$ and $t_C$
correspond to the times when the signal detected respectively by
coils A, B and C is maximum. The thermal dependence of this
velocity is shown in the inset of Fig. \ref{fig:coils}, according
to which the average value of $v$ is about 20 m/s at 3.5 K. As $T$
and $x$ are strongly correlated, the velocity of the avalanche
also depends on the value of $x$ when the abrupt phase transition
is ignited.

\begin{figure}
\includegraphics[width=\columnwidth]{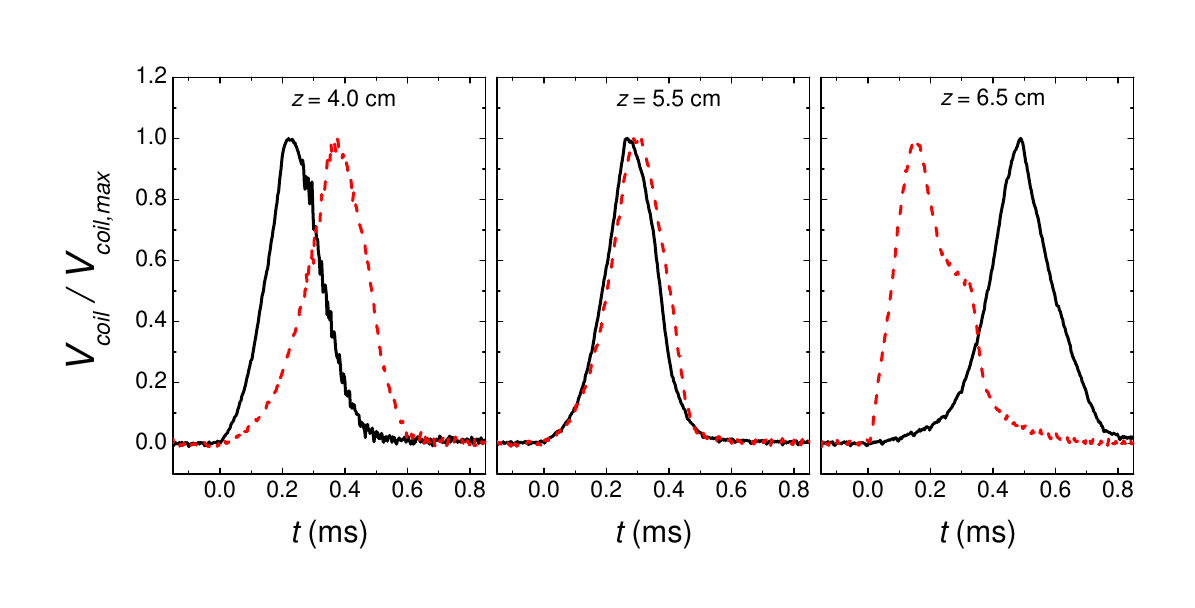}
\caption{Time evolution of the voltages detected by coils 1 (solid
line) and 2 (dashed line) for different locations of the sample.
From left to right, $z=4.0$ cm (left panel), $z=5.5$ cm (central
panel), and $z=6.5$ cm (right panel). All signals have been
normalized to their maximum values. As $z$ goes up and the
gradient of the applied magnetic field increases, the time
sequence of the signal registered by the two detection coils is
reversed. The experiment was performed at 3 K.}
\label{fig:gradiente1}
\end{figure}

\begin{figure}
\includegraphics[width=\columnwidth]{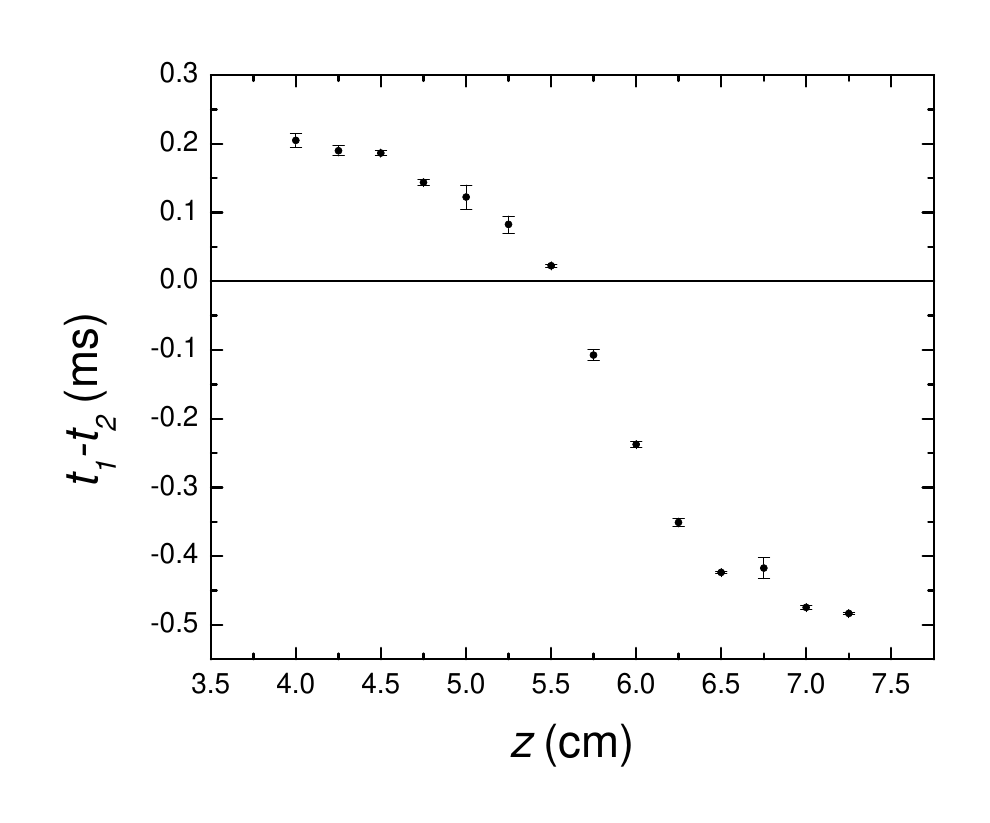}
\caption{Variation of the time difference between the maxima of
the signal of the two coils, $t_1-t_2$, with respect to $z$. The
avalanches were ignited at 3.5 K.} \label{fig:gradiente2}
\end{figure}

The results shown in Fig. \ref{fig:coils} correspond to the case
when the applied magnetic field is uniform along the sample. We
have also performed experiments shifting the position of the
sample along the $z$ axis. In these cases the sample suffers the
effect of the field and its gradient in such a way that we can
obtain variations of the magnetic field along the longest
dimension up to a 10\% of the maximum value of $H$ applied. In
this case we have used two new detection coils, named 1 and 2,
located at the positions where coils A and C previously were.
Figure \ref{fig:gradiente1} shows the voltages detected by these
coils as a function of time for three different positions along
the $z$ direction. Considering that the variation of magnetic
field between the two edges of the polycrystal can be roughly
estimated as $\Delta H \sim H_{max}\cdot$d$h/$d$z\cdot L$, where
$L$ is the length of the sample, $H_{max}\sim 28$ kOe and taking
d$h$/d$z$ from Fig. \ref{fig:setup}b, values of $\Delta H\sim
100$, 800, and 2000 Oe can be obtained at $z=4.0$, 5.5 and 6.5 cm.
The time sequence of the peaks measured by the detection coils at
Fig. \ref{fig:gradiente1} shows that the presence of a field
gradient affects dramatically the origin and propagation of the
abrupt phase transition along the sample. This is more clearly
seen at Fig. \ref{fig:gradiente2}, where we have plotted the time
difference between the maxima of the signals in the two coils,
$t_1-t_2$, as a function of $z$. The sign change in this magnitude
reflects that, as the strength of the field gradient increases,
the local magnetic field at different points in the sample varies
and thus the distribution of energy barriers $U(H)$ between the
AFM and FM phases along the sample is different from that in the
case of a homogeneous applied magnetic field. As a consequence,
the place of the sample where the abrupt phase transition begins
must shift, leading to changes in the signal measured by the
detection coils.

In conclusion we have demonstrated that the magnetic avalanches
associated to the transition from the CO-AFM metastable state into
the CD-FM metallic state by the influence of the applied magnetic
field correspond to a phase transformation mechanism that moves
along the sample, being ignited in the inside and propagating to
the edges. This mechanism reflects, at macroscopic scales, the
same physics showed\cite{Wu} at microscopic ones: when the applied
field overcomes certain temperature- and state-dependent threshold
value the system becomes unblocked, inducing the growth of the
CD-FM phase against the unstable CO-AFM phase. We have also shown
that by applying a magnetic field gradient it is possible to
change the place in the sample where this threshold is overcome,
allowing to modify the position where the avalanche is ignited.

This work was supported by contract MAT2005-06162 from the Spanish
Ministerio de Educación y Ciencia (MEyC). A. H.-M. and F. M. thank
the MEyC for a research grant. J. M. H. thanks the MEyC and
Universitat de Barcelona for a Ramón y Cajal research contract.
The authors thank Gabriela Leyva for preparing the sample.

\end{document}